\documentclass[iop]{emulateapj}

\newcommand{\flux}{\,erg\,s$^{-1}$\,cm$^{-2}$}

\newcommand{\cm}{\,cm$^{-2}$}
\newcommand{\nh}{$N_\mathrm{H}$}
\def\etal{{et al.}}

\shorttitle{A Radio-bright Binary in M10}
\shortauthors{Shishkovsky et al.}

\begin{document}

\title{The MAVERIC Survey: A Red Straggler Binary with an Invisible Companion in the Galactic Globular Cluster M10}

\author{Laura Shishkovsky\altaffilmark{1}, Jay Strader\altaffilmark{1}, Laura Chomiuk\altaffilmark{1}, Arash Bahramian\altaffilmark{1}, 
Evangelia Tremou\altaffilmark{1}, Kwan-Lok Li\altaffilmark{1}, Ricardo Salinas\altaffilmark{2}, Vlad Tudor\altaffilmark{3}, James C.A.~Miller-Jones\altaffilmark{3}, Thomas J.~Maccarone\altaffilmark{4}, Craig O.~Heinke\altaffilmark{5}, Gregory R.~Sivakoff\altaffilmark{5}}

\altaffiltext{1}{Center for Data Intensive and Time Domain Astronomy, Department of Physics and Astronomy, Michigan State University, East Lansing MI, USA;  \email{shishko1@msu.edu}
}
\altaffiltext{2}{Gemini Observatory}
\altaffiltext{3}{International Centre for Radio Astronomy Research -- Curtin University, GPO Box U1987, Perth, WA 6845, Australia}
\altaffiltext{4}{Texas Tech University}
\altaffiltext{5}{University of Alberta}

\begin{abstract}
We present the discovery and characterization of a radio-bright binary in the Galactic globular cluster M10. First identified in deep radio continuum data from the Karl G. Jansky Very Large Array, M10-VLA1 has a  flux density of $27\pm4$ $\mu$Jy at 7.4 GHz and a flat to inverted radio spectrum. \emph{Chandra} imaging shows an X-ray source with $L_X \approx 10^{31}$ erg s$^{-1}$ matching the location of the radio source. This places M10-VLA1 within the scatter of the radio--X-ray luminosity correlation for quiescent stellar-mass black holes, and a black hole X-ray binary is a viable explanation for this system. The radio and X-ray properties of the source disfavor, though do not rule out, identification as an accreting neutron star or white dwarf system. Optical imaging from the \emph{Hubble Space Telescope} and spectroscopy from the SOAR telescope show the system has an orbital period of 3.339 d and an unusual ``red straggler" component: an evolved star found redward of M10's red giant branch. These data also show UV/optical variability and double-peaked H$\alpha$ emission characteristic of an accretion disk. However, SOAR spectroscopic monitoring reveals that the velocity semi-amplitude of the red straggler is low. We conclude that M10-VLA1 is most likely either a quiescent black hole X-ray binary with a rather face-on ($ i < $ 4$^{\circ}$) orientation or an unusual flaring RS CVn-type active binary, and discuss future observations that could distinguish between these possibilities.

\end{abstract}


\section{Introduction}

Due to their high stellar densities and large populations of stellar remnants, globular clusters are unique environments for the efficient formation of binaries with compact objects. The pathways to forming these low-mass X-ray binaries in globular clusters include tidal capture, three-body binary exchange, and direct stellar collisions with compact objects (e.g., Fabian, Pringle \& Rees 1975; Hills 1976; Verbunt \& Hut 1983; Bailyn \& Grindlay 1990;  Davies \& Hansen 1998; Ivanova et al.~2008; Ivanova et al.~2010)---in contrast with field X-ray binaries, which likely evolved as isolated systems. The formation of low-mass X-ray binaries through close encounters accounts for the high specific abundance of X-ray binaries in globular clusters in both the Milky Way and in other galaxies (e.g., Pooley et al.~2003; Kundu et al.~2002). 

While a substantial fraction of field low-mass X-ray binaries in the Milky Way host black holes (e.g., Tetarenko et al. 2016a; Remillard \& McClintock 2006), the overwhelming majority of  low-mass X-ray binaries in globular clusters---at least those which are bright and well-studied---host neutron stars rather than black holes (Verbunt \& Lewin 2006; Bahramian et al.~2014), typically identified through Type I X-ray bursts (Lewin \etal~1993).

Many authors have argued that the relative paucity of black holes was real, beginning with analytic arguments about the fate of black holes in the dense cluster environment. After formation, any black holes that do not receive strong natal kicks will sink to center of the cluster and become segregated from less massive stars. In these close quarters, the black holes will form tight binaries that are largely ejected, through interactions with other black holes or black hole--black hole binaries. This process was argued to continue until all (or nearly all) black holes were depleted from the cluster (Sigurdsson \& Hernquist 1993; Kulkarni et al.~1993).

Parallel observational and theoretical tracks have led to a reconsideration of this conclusion. Several globular clusters in external galaxies may contain black holes accreting near the Eddington luminosity, with the quality of the evidence ranging from suggestive to compelling (e.g., Sarazin, Irwin \& Bregman 2001; Maccarone et al.~2007; Zepf et al.~2008; Irwin et al.~2010; Peacock et al.~2012). In the Milky Way, low-luminosity black hole candidates have been identified by a combination of radio continuum, X-ray, and optical data in the globular clusters M22 (Strader et al.~2012), M62 (Chomiuk et al.~2013), and 47 Tuc (Miller-Jones et al.~2015; Bahramian et al.~2017). A number of theoretical papers have concluded that black hole ejection is less efficient than originally thought, since a putative subcluster of black holes cannot remain dynamically isolated from the rest of the cluster as its mass declines (Mackey et al. 2008; Moody \& Sigurdsson 2009; Sippel \& Hurley 2013; Morscher et al.~2013; Breen \& Heggie 2013; Heggie \& Giersz 2014; Morscher et al.~2015). This work has accelerated since the discovery of merging black hole--black hole binaries by Advanced LIGO (Abbott et al.~2016), and the dynamical formation of black hole--black hole binaries in globular clusters may be an important or even dominant channel for such systems (Rodriguez et al.~2016; Chatterjee et al.~2017).

No bright ($>$ $10^{36}$ erg s$^{-1}$) X-ray binary in a Galactic globular cluster has ever been identified to host a black hole; the candidates identified thus far are all in quiescence. Given the limited number of bright X-ray binaries in clusters, this could be due to small number statistics or could reflect unusual formation channels for cluster X-ray binaries compared to field systems. For example, short-period black hole X-ray binaries could undergo shorter, less-luminous outbursts that would not be detected by all-sky X-ray monitors (Maccarone \& Patruno 2013; Knevitt et al.~2014).

In any case, black hole low-mass X-ray binaries are expected to spend most of their lives in a low-luminosity state with  $L_X \sim 10^{30}$--$10^{33}$ erg s$^{-1}$
(Corbel et al.~2006) In this state it is typically not possible to separate them from other X-ray sources, such as compact binaries containing white dwarfs or neutron stars, or even active binaries, on the basis of X-ray observations alone. However, in quiescence, black holes are observed to emit steady flat-spectrum radio continuum emission, thought to originate via partially self-absorbed synchrotron radiation from compact jets (Blandford \& K{\"o}nigl 1979). 

The possibility of identifying quiescent black hole low-mass X-ray binaries through radio continuum emission motivated our group to conduct a systematic survey of 50 Galactic globular clusters using radio continuum observations from the upgraded Karl G.~Jansky Very Large Array (VLA) and the Australia Telescope Compact Array (ATCA). We name this survey MAVERIC (Milky-way ATCA and VLA Exploration of Radio-sources In Clusters).

Here we present a multi-wavelength study of a radio-selected black hole candidate in the Galactic globular cluster M10 (NGC 6254; D = 4.4 kpc; Hurley et al.~1989; Harris 1996 (2010 edition)). M10 has a [Fe/H] $\sim -1.5$, and its mass is about 1.5 $\times {10^5}$ $ {M_{\odot}}$ (McLaughlin \& van der Marel 2005; Haynes et al.~2008). In Section 2, we discuss our VLA observations, \textit{Chandra} X-ray data, \emph{Hubble Space Telescope} optical photometry, and ground-based SOAR spectroscopy of the system. In Section 3 we discuss the properties of the binary: identity of the binary companion, orbital parameters, and mass constraints. In Section 4 we discuss the interpretation of the system, and summarize our findings and discuss future work in Section 5.

\section{Observations and Analysis}

\subsection{Radio}

We observed M10 using the VLA in early 2014 in five 2-hr blocks (10 hr total, about 7 hr on source). The observations were done in A configuration and with C band receivers, with two 1-GHz basebands centered at 5.0 GHz and 7.4 GHz. Each baseband consisted of eight spectral windows, each 128 MHz wide, sampled with 64 channels of width 2 MHz. We used 8-bit samplers and obtained full polarization products.

3C286 was used as a flux density and bandpass calibrator, while J1651+0129 was used as a phase calibrator. The radio data from each epoch were reduced using {\it Common Astronomy Software Application} (CASA) (McMullin et al. 2007) version 4.2.2 with version 1.3.1 of the VLA calibration pipeline.\footnote{\url{https://science.nrao.edu/facilities/vla/data-processing/pipeline}} The pipeline imports raw data, applies preliminary flags, and then iteratively calibrates the data while running automatic flagging algorithms for radio frequency interference. We manually flagged any remaining corrupt data and then re-ran the pipeline. Once the target was properly calibrated and flagged, we imaged each baseband separately. 
The field of view was selected to match the FWHM size of the primary beam: 11{\arcmin} at 5.0 GHz and 7.5{\arcmin} at  7.4 GHz. We used Briggs weighting with a robust parameter of 1, and nterms = 2 to account for the non-zero spectral indices of the various sources in the field. The synthesized beams for the 5.0 and 7.4 GHz images are $0.75\arcsec \times 0.36\arcsec$
and $0.53\arcsec \times 0.26\arcsec$, respectively. 

In a future paper we will discuss the details of all radio continuum sources detected within the VLA's primary beam. This paper focuses on the only source within the 48\arcsec\ (1.0 pc; Dalessandro et al. 2011) cluster core radius  detected in both the upper and lower basebands. This source, which we term M10-VLA1, is located at a J2000 position of (R.A., Dec.) = (16:57:08.478 {$\pm$} 0.013s, $-$04:05:55.72 {$\pm$} 0.19\arcsec), only 10\arcsec\ (0.2 pc) in projection from the photometric center of the cluster (Goldsbury et al.~2010). The source was clearly detected in both basebands during the first 2-hr observing block (on 2014 Feb 20; Table 1) and marginally detected at 5 GHz on 2014 Apr 29; it was not detected in the other three blocks in either baseband. The images of the source from the first block are shown in Figure 1.

\begin{figure}[!tbp]
  \centering
  {\includegraphics[width=0.4\textwidth]{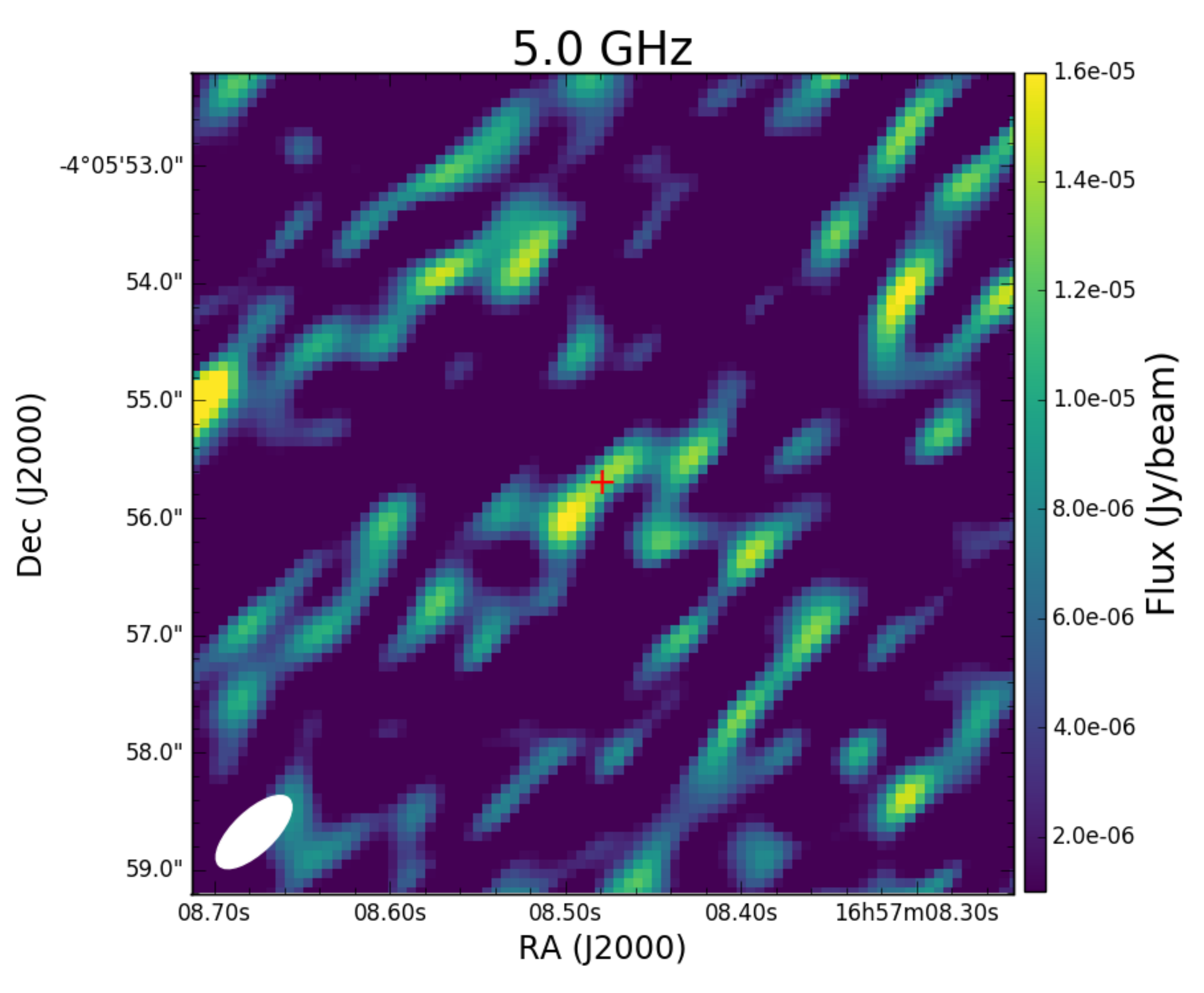}\label{fig:f1}}
  \hfill
  
 { \includegraphics[width=0.4\textwidth]{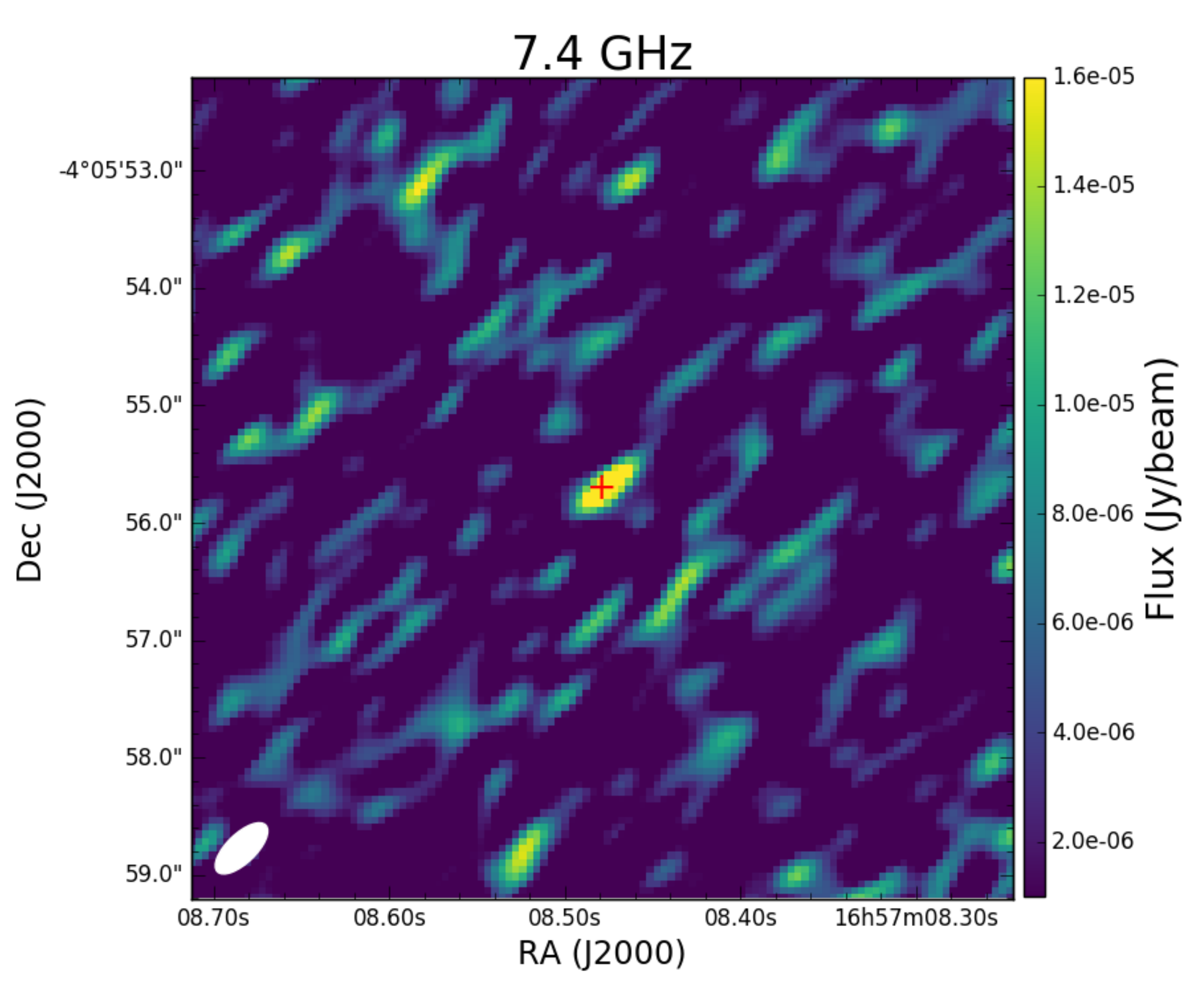}\label{fig:f2}}
  \caption{VLA 5.0 GHz (top) and 7.4 GHz (bottom) radio images of the 2014 Feb 20 detection of M10-VLA1. The red cross marks the location of the source center given by {\tt imfit} from the combined 6.0 GHz image of the 2014 Feb 20 observation. We note that the source position is dominated by the 7.4 GHz flux --- at 5.0 GHz the detection is just 3$\sigma$. The image synthesized beam is denoted as a white ellipse in the bottom left corner of each image. 
}
\end{figure}

We determined the flux densities of M10-VLA1 by fitting a point source in the image plane using the task {\tt imfit} in CASA, constraining the source size to the dimensions of the synthesized beam. On 2014 Feb 20, the flux density was $16.2\pm5.4$ $\mu$Jy (5.0 GHz) and $27.2\pm4.2$ $\mu$Jy (7.4 GHz, giving a luminosity spectral density of $6 \times 10^{26}$ erg s$^{-1}$ GHz$^{-1}$). Assuming a power-law frequency dependence ($S_\nu \propto {\nu}^{\alpha}$) the source has evidence for a flat to inverted spectrum, with $\alpha = 1.3\pm0.9$. There was no clear detection of M10-VLA1 in either baseband for the subsequent epochs, taken 36 to 68 days after the initial data. There is a marginal 5 GHz detection of the source on 2014 Apr 29 ($14.7\pm4.5 \mu$Jy beam$^{-1}$), accompanied by a non-detection at 7.4 GHz at this epoch ($3\sigma$ upper limit of $< 10.8 \mu$Jy beam$^{-1}$). This suggests a spectral index of $\alpha < -0.8$, which is much steeper than the 2014 Feb 20 epoch. However, given the large uncertainties in the flux densities, an inverted spectral slope consistent with the earlier epoch ($\alpha = 1.3\pm0.9$) cannot be ruled out at even the 2$\sigma$ level. The flux densities and 3{$\sigma$} upper limits from the initial detection and the other epochs are listed in Table 1. The individual epochs with non-detections were also combined and imaged, yielding no detections and 3{$\sigma$} upper limits of 7.7$\mu$Jy beam$^{-1}$ (5.0 GHz) and 5.8$\mu$Jy beam$^{-1}$ (7.4 GHz).

\begin{deluxetable*}{llccccc}
\centering
\tabletypesize{\scriptsize}
\tablecolumns{10} 
\tablewidth{0pt} 

\tablecaption{VLA Radio Flux Densities of M10-VLA1 \label{tab:vla}}
\tablehead{
\colhead{Epoch Date}           &
\colhead{Epoch Date}           &
\colhead{Time}           &  
\colhead{ 5.0 GHz Peak Flux} &
\colhead{RMS} &
\colhead{ 7.4 GHz Peak Flux} &
\colhead{RMS} 
 \\
\colhead{} &
\colhead{(MJD)} &
\colhead{(UTC)}           &
\colhead{($\mu$Jy)}&
\colhead{($\mu$Jy/beam)}&
\colhead{($\mu$Jy)}&
\colhead{($\mu$Jy/beam)}
}
\startdata

Feb 20, 2014    &56708.4      &09:19:32 $-$ 11:19:12     &   16.2	&5.4	      &27.2 &4.2        \\    
Mar 28, 2014    &56744.3      &07:32:59 $-$ 09:32:36      &   {$<$} 13.8 	&4.6	 &{$<$} 11.4  &3.8	    \\ 
Apr 7, 2014     &56754.3      &06:49:09 $-$ 08:48:46       &   {$<$}  14.7	&4.9	 &{$<$} 11.4  &3.8        \\  
Apr 10, 2014    &56757.4      &09:07:49 $-$ 08:06:32      &   {$<$} 15.9	&5.3      &{$<$} 12.3  &4.1        \\ 
Apr 29, 2014    &56776.2      &04:56:49 $-$ 06:56:29         &   14.7 &4.5      &{$<$} 10.8  &3.6        \\
\enddata
\tablecomments{The flux density upper limits represent $3\sigma$ limits.}
\end{deluxetable*}

To check for short-term variability, we re-imaged the 2014 Feb 20 epoch on timescales of about 10 min (averaging 9 target scans, each of 62.8 s duration) in each frequency band. For
both basebands, the  flux densities were constant within the uncertainties across the observation. Therefore, there is no evidence that M10-VLA1 was variable over the 2 hr observation on 2014 Feb 20.

\subsection{X-ray}
\subsubsection{Chandra}

Subsequent to the detection of the radio continuum source, we observed M10 in X-rays with \textit{Chandra}/ACIS-S for 32.6~ksec on 2015 May 08.  The \emph{Chandra} image is shown in Figure 2. We used \texttt{CIAO} 4.7 and \texttt{CalDB} 4.6.9 to complete the \emph{Chandra} data analysis (Fruscione et al. 2006). Using the \texttt{CIAO} \texttt{wavdetect} task for source detection, a faint X-ray source with $\sim 12$ net counts is clearly present at the position of the radio source to within the \textit{Chandra} absolute astrometric accuracy of 0.6\arcsec. There is no evidence for variability, however the low number of counts prevents us from concluding this definitively. The field is not crowded in the X-rays: only 10 X-ray sources are detected by \texttt{wavdetect} within the 1.95\arcmin\ half-light radius of M10 (equivalent to a source density of $\sim 2\times10^{-4}$ arcsec$^{-2}$). There are less than ten 5$\sigma$ radio sources (at 6.0 GHz) within the half-light radius, suggesting extremely low odds of a chance match between an X-ray source and the radio source. Therefore the X-ray source is almost certainly associated with the radio source.

Using a circular source region of 2\arcsec\ radius and a source-free annulus background around the source (inner/outer radii: 10{\arcsec}/20{\arcsec}), we extracted the X-ray spectrum with \texttt{specextract}. The spectral analysis was performed using XSPEC (version 12.9.0; Arnaud 1996), assuming Anders \& Grevesse (1989) abundances and Verner et al. (1996) absorption cross-sections. Given the limited number of photons, we analyzed the spectrum by binning the data with at least one count per bin and using the XSPEC operation cstat, a modified version of the Cash statistic (Cash 1979) for studying low-count spectra.\footnote{\url{https://heasarc.gsfc.nasa.gov/xanadu/xspec/manual/~XSappendixStatistics.html}} Assuming an absorbed power-law with an X-ray absorption of \nh$\,=2.44\times10^{21}$\cm\ (frozen to this value, derived from $E(B-V) = 0.25$--0.28; Schlafly \& Finkbeiner 2011; Harris 1996; Bahramian et al.~2015; Foight et al.~2016), the best-fit photon index is $\Gamma=2.8^{+1.1}_{-1.0}$ and the absorption-corrected 0.5--10 keV flux is $4.5^{+2.5}_{-1.8}\times10^{-15}$\flux\ (all uncertainties represent 90\% confidence intervals). This flux is equivalent to $L_X \sim 1.0^{+0.6}_{-0.4} \times 10^{31}$ erg s$^{-1}$ at the distance of M10. The spectrum was also fit using a MEKAL model (Mewe, Gronenschild \& van den Oord 1985; Mewe, Lemen \& van den Oord 1986; Kaastra 1992; Liedahl, Osterheld \& Goldstein 1995), also with the cstat operation. Again we assume \nh$\,=2.44\times10^{21}$\cm\, and that the metallicity is 0.03 solar. The best-fit temperature from the MEKAL model is $1.1^{+2.1}_{-0.5}$ keV, and the unabsorbed flux from 0.5 - 10 keV is $3.6^{+2.1}_{-1.4}\times10^{-15}$\flux at 90\% confidence. This flux corresponds to $L_X \sim 8.4^{+4.9}_{-3.2} \times 10^{30}$ erg s$^{-1}$ at the distance of M10.

\begin{figure}[t]
  \centering 
  {\includegraphics[width=0.45\textwidth]{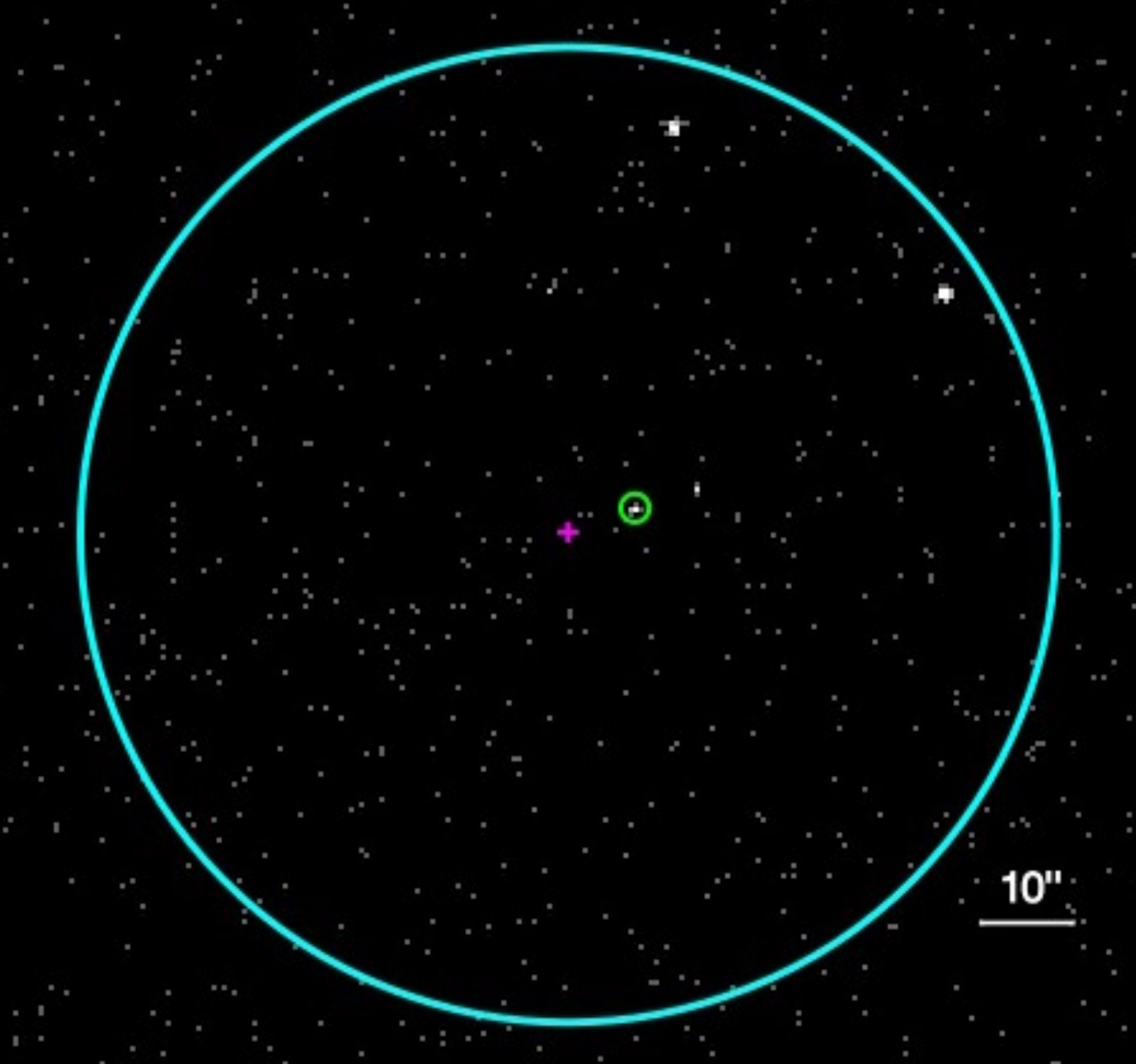}\label{fig:chandra}}
  \hfill
  \caption{{\it{Chandra}}/ACIS-S X-ray image of the core of M10 (blue circle; radius $46.2\arcsec \sim 1.0$ pc) in the 0.3 - 7.0 keV energy band. The magenta cross marks the cluster photometric center (Goldsbury et al. 2010). The X-ray source associated with M10-VLA1 is circled in green.}
\end{figure}

\subsubsection{Swift}

There are several \emph{Swift}/XRT observations of M10, including one taken essentially simultaneously (within one day) of the initial radio observations reported in \S 2.1. These X-ray observations are detailed in Table 2. M10-VLA1 is not detected in any of these observations. To determine flux upper limits, we assumed the best-fit parameters from the \emph{Chandra} observations, using an extraction radius of 25\arcsec\ to reduce the chance of contamination from the nearby X-ray sources. These  background-subtracted upper limits are at the 95\% confidence level and assume an energy range of 0.5--10 keV. Table 2 also contains a stacked upper limit from a combination of all the \emph{Swift} observations. 

The individual epoch upper limits are all in the range of $<$ (5--16) $\times 10^{31}$ erg s$^{-1}$, and the stacked upper limit is $< 2.2  \times 10^{31}$ erg s$^{-1}$, hence consistent with the \emph{Chandra} flux value. 

\begin{deluxetable}{llcc}
\centering

\tablecaption{Swift X-ray Constraints \label{tab:swift}}
\tablehead{
\colhead{Epoch Date}&
\colhead{Epoch Date}&
\colhead{Effective time} &
\colhead{Luminosity limit} 
 \\
\colhead{}&
\colhead{(MJD)} &
\colhead{(s)}&
\colhead{(erg s$^{-1}$)}
}
\startdata
Oct 30, 2009 & 55117.81375 & 1928 &  $< 4.7\times10^{31}$ \\
Feb 21, 2014 & 56709.72004 & 1071 & $<5.3\times10^{31}$ \\
Oct 21, 2014 & 56951.26891 & 402 & $<1.6\times10^{32}$  \\
Jan 20, 2015 & 57042.62936 & 465 & $<1.3\times10^{32}$ \\
Oct 20, 2015 & 57315.31564 & 787 & $<7.7\times10^{31}$  \\
Jan 20, 2016 & 57407.33950 & 1631 & $<5.6\times10^{31}$  \\
Jan 21, 2016 & 57408.13957 & 989 & $<1.3\times10^{32}$ \\
\hline
Combined &  & 7272 & $<2.2\times10^{31}$  \\
\enddata
\tablecomments{All limits are at the 95\% level and over the energy range 0.5--10 keV. Luminosities assume a distance of 4.4 kpc.}
\end{deluxetable}

\subsection{Optical Photometry}

\emph{Hubble Space Telescope}/Advanced Camera for Surveys (ACS) photometry for M10 in $F606W$ and $F814W$ has already been published as part of the ACS survey of Galactic globular clusters (Sarajedini et al.~2007; Anderson et al.~2008).
We corrected the astrometry of these images using a large number of \emph{Gaia} stars (Gaia collaboration et al.~2016), finding an rms of about $0.02\arcsec$ per coordinate. The closest optical source to M10-VLA1 is located $0.116\arcsec$ from the radio source, consistent within the combined positional uncertainties of the radio and optical sources (Figure 3). The unusual optical properties of this source (see below) confirm its identity as the optical counterpart to M10-VLA1.

\begin{figure}[t]
  \centering
  {\includegraphics[width=0.425\textwidth]{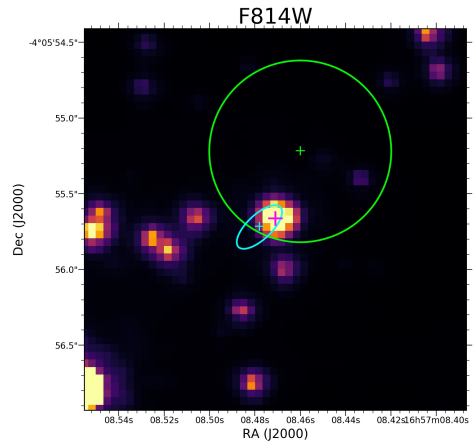}\label{fig:hst}}
  \hfill
  \caption{{\it{HST}} $F814W$ image of the optical counterpart to M10-VLA1 with the detection positions of the {\it{Chandra}} observation and 20 Feb 2014 VLA observation overlaid. The optical counterpart is marked by the magenta cross. The VLA position and associated positional error is shown by the blue cross and blue ellipse. The {\it{Chandra}} position is shown by the green cross, with the green circle representing the {\it{Chandra}} astrometric accuracy.}
\end{figure}

Figure 4 shows the position of the source in a color-magnitude diagram (CMD) of M10. Here the plotted stars are restricted to a radius of 15\arcsec\ around M10-VLA1 to reduce the effects of differential reddening on the distribution of stars in the CMD (we note all photometry listed is as observed, not corrected for the substantial foreground reddening ($E(B-V) = 0.25$--0.28)). The optical counterpart to M10-VLA1 has $F606W = 17.238\pm0.005$ and $F606W$-$F814W$ = $1.036\pm0.009$ mag on the VEGAMAG system. It sits $\sim 0.19$ mag redward  in $F606W-F814W$ of the lower giant branch members of M10 of the same $F606W$ mag. The unusual color would normally lead to the conclusion that the star is not a cluster member. However, the spectroscopic observations (Section \ref{optspec}) show the star has a radial velocity consistent with the cluster systemic velocity, and it is located only 0.2 pc in projection from the center of M10. Hence we conclude that the star is indeed a member of the cluster but with an unexpectedly low effective temperature for its luminosity. The nomenclature of these stars is somewhat confusing, and we follow the recent suggestion of Geller et al.~(2017) that such stars, when brighter than the subgiant branch, be referred to as ``red stragglers"; the term ``sub-subgiants" is reserved for systems fainter than normal subgiants.

Using the solar-scaled MIST isochrones (Dotter 2016; Choi et al.~2016) for [$Z$/H] = --1.2, an age of 12 Gyr, and assumed $E(B-V) = 0.28$, giants with $F606W-F814W$ matching that observed for M10-VLA1 have $T_{eff} \sim 4800$ K (of course, these giants are much more luminous than M10-VLA1). The bolometric luminosity inferred from the temperature and the $F814W$ magnitude is about $4.4 L_{\odot}$.

While these ACS data were obtained on 2006 Mar 5, the ground-based SOAR telescope photometry of Salinas et al.~(2016), taken in July 2015, show a qualitatively similar location of the star in a $g-i$ vs.~$i$ CMD (Figure 5). This implies the location of M10-VLA1 in the ACS CMD is not a fluke, but is persistent over timescales of years. The star's short-term variability is poorly constrained: the SOAR photometry covered about 6.7 hr, over which the star became $\sim 0.01$ mag fainter in $i$, but these data cover only a tiny fraction of the 80 hr orbital period (see \S \ref{optspec}).

Additional HST/WFC3 imaging in $F275W$, $F336W$, and $F438W$ was obtained on 2013 Aug 16 and 2014 May 27 with Wide Field Camera 3 (WFC3). The time-averaged photometry at these bands is available from the catalogs of Soto et al.~(2017), which represent preliminary measurements from the \emph{HST} Treasury program of Piotto et al.~(2015). These values are $F275W = 20.80\pm0.18$, $F336W = 19.09\pm0.14$, and $F438W = 18.80\pm0.09$. As the uncertainties in these measurements are substantially larger than for other stars of this brightness, we individually photometered the two epochs to search for variability. We found that the source brightened by $-0.21\pm0.03$ mag in $F275W$, $-0.16\pm0.03$ mag in $F336W$, and $-0.23\pm0.02$ mag in $F438W$ between the two epochs. A color-magnitude diagram with the UV photometry, constructed in an identical way to the optical color-magnitude diagram, is given in Figure 4.

Given this evidence for variability, we cannot combine the data for these bluer bands with the $F606W$ and $F814W$ data to model the spectral energy distribution. CMDs made with the bluer filters confirm that the star is an outlier, sitting far redward of the main locus of stars in any pair of filters, consistent with the unusual $F606W$-$F814W$ color. 
If we use the same MIST isochrones discussed above to infer the $T_{eff}$ from $F275W$--$F438W$ or $F336W$--$F438W$,
we find $T_{eff} \sim 5100$ K for both, warmer than found for the $F606W$-$F814W$ color. This difference could be due to a real change in the disk-averaged $T_{eff}$ (e.g., from starspots) or due to a varying contribution from a warmer component, such as an accretion disk. Future simultaneous photometry over a broad baseline could allow one to better constrain the presence of a hot companion or a disk.

\begin{figure}[!tbp]
  \centering
  {\includegraphics[width=0.4\textwidth]{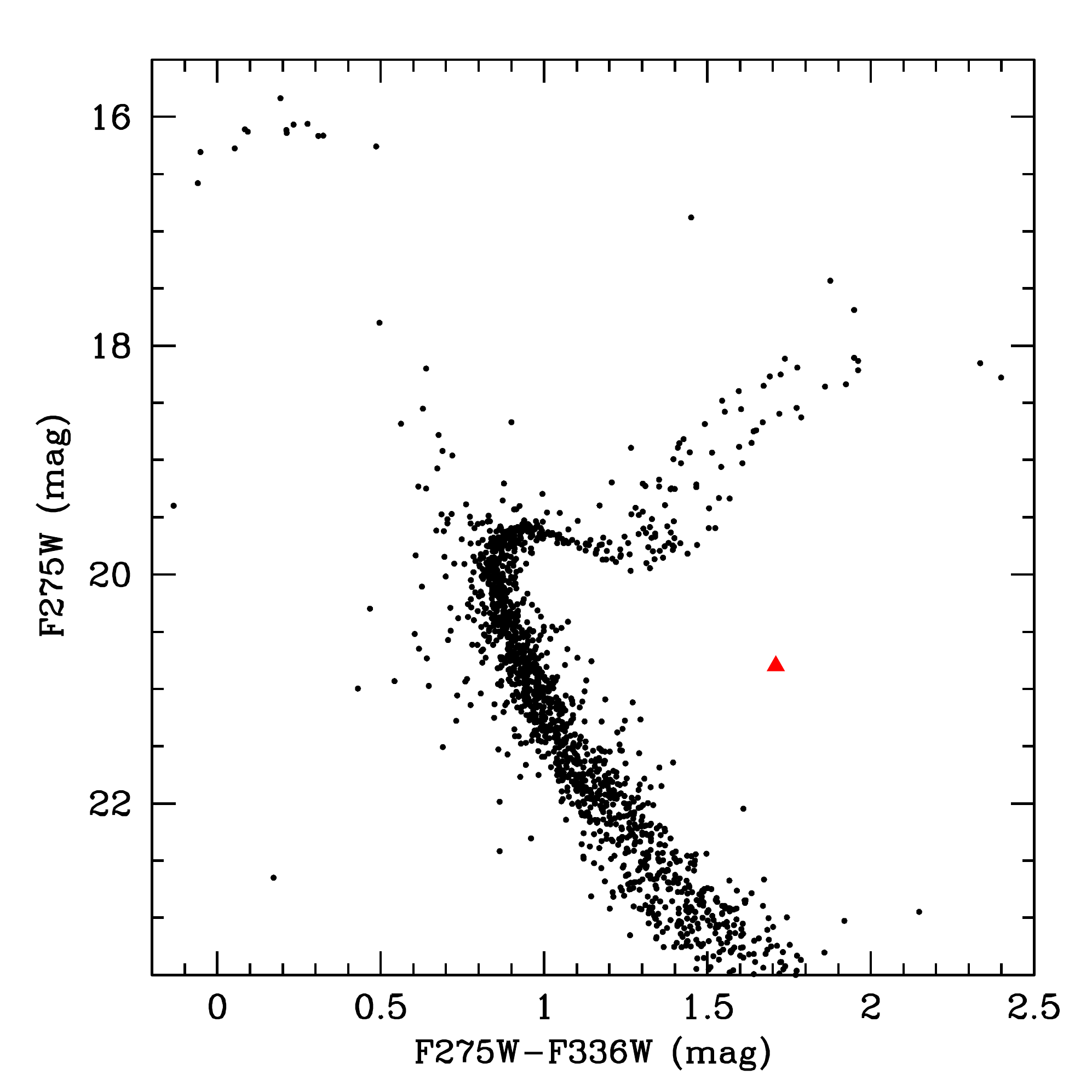}\label{fig:f1}}
  \hfill
  
 { \includegraphics[width=0.4\textwidth]{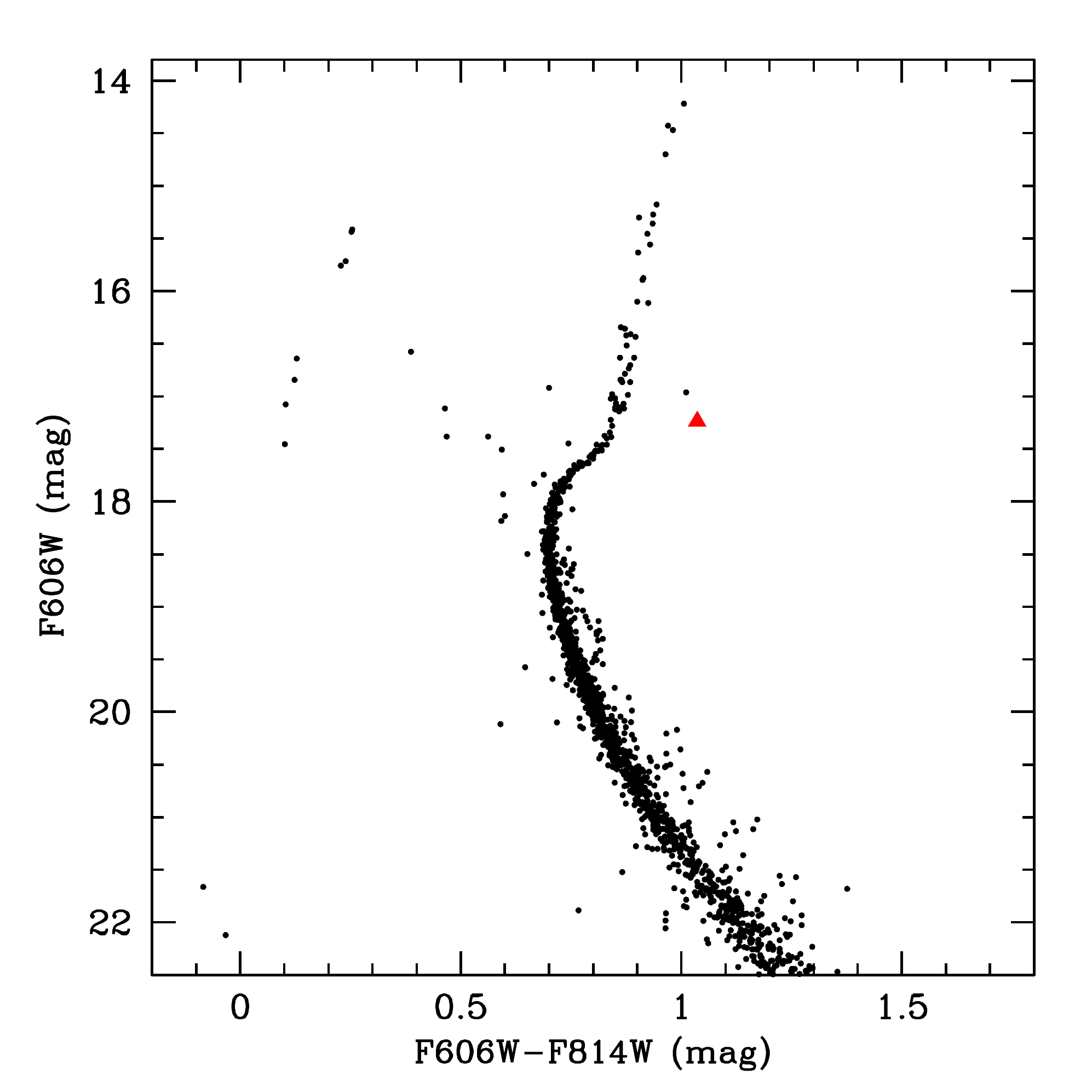}\label{fig:f2}}
  \caption{{\it{HST}} color-magnitude diagrams in $F275W$ vs. $F275W-F336W$ (top) and $F606W$ vs. $F606W-F814W$ (bottom) of the stars within 15{\arcsec} of M10-VLA1. The optical counterpart to M10-VLA1 is shown with the red triangle.}
\end{figure}

\begin{figure}[t]
  \centering
  {\includegraphics[width=0.45\textwidth]{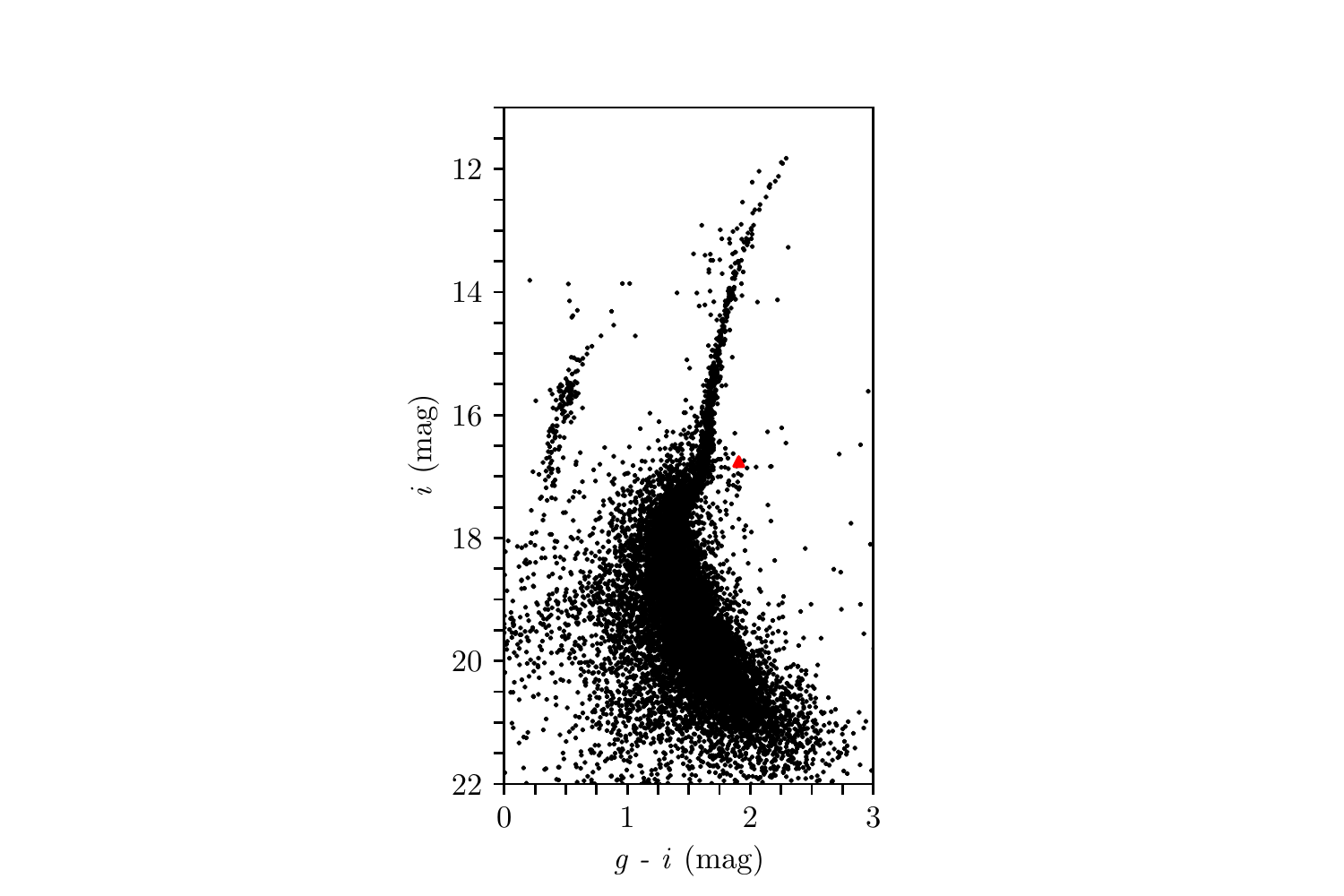}\label{fig:soarcmd}}
  \hfill
  \caption{{\it{SOAR}} color-magnitude diagram in $i$ vs. $g-i$ of M10, observed $\sim$9 years after the \emph{HST} photometry presented in Figure 3. The optical counterpart to M10-VLA1 is shown with the red triangle. The similarity to its location in the Figure 3 CMD shows that its unusual color is persistent over long timescales and is not due to variability.}
\end{figure}

\subsection{Optical Spectroscopy}\label{optspec}

We initiated  spectroscopy of the optical counterpart to M10-VLA1 in 2015, using the Goodman spectrograph on the SOAR 4.1-m telescope (Clemens et al.~2004). All observations were made with a 0.95\arcsec\ slit, but used several different gratings: some with a 1200 l mm$^{-1}$ grating (resolution 1.7 \AA; range 5380--6640 \AA, for studying H$\alpha$), and radial velocity measurements with a 2100 l mm$^{-1}$ grating (resolution 0.9 \AA; range 5020--5660 \AA) or a  2400 l mm$^{-1}$ grating (resolution 0.7 \AA; range 5080--5610 \AA). Typical individual exposure times were 900 sec each, with two spectra generally taken back-to-back on a particular night. Spectra were reduced and wavelength calibrated with a FeAr arc lamp using standard routines in IRAF. The spectra covering H$\alpha$ show clear H$\alpha$ emission (Figure 7), indicative of binary interaction, providing additional evidence that this object is the counterpart to the radio source.

We determined radial velocities through cross-correlation over the region 5150--5300 \AA\ with spectral templates of similar spectral type taken with the same setup. Given the long period of the system (see below), we used a weighted average of the radial velocities for the consecutive 900-sec spectra to represent each epoch of data. The barycentric radial velocities are listed in Table 3. The corresponding observation times are given as Barycentric Julian Dates on the TDB system (Eastman et al.~2010).

\section{Binary Properties and Analysis}

\subsection{Orbital Parameters}

Using the custom Keplerian sampler \emph{TheJoker}\footnote{https://github.com/adrn/thejoker/} (Price-Whelan et al.~2017), we initially fit a circular model to the 20 radial velocity epochs. The posterior distributions for the fitted parameters were all unimodal and close to Gaussian, with median values: period $P = 3.3391\pm0.0010$ d, systemic velocity $v_{sys} = 86.3\pm1.1$ km s$^{-1}$, semi-amplitude $K_2 = 12.5\pm1.5$ km s$^{-1}$, and the ascending node of the compact object $T_0 = 2457169.1292\pm0.0846$ d. A fit with these values is plotted in Figure 6. The residuals around the best orbital fit have an rms dispersion of 3.9 km s$^{-1}$.

\begin{figure}[t]
  \centering
  
  {\includegraphics[width=0.425\textwidth]{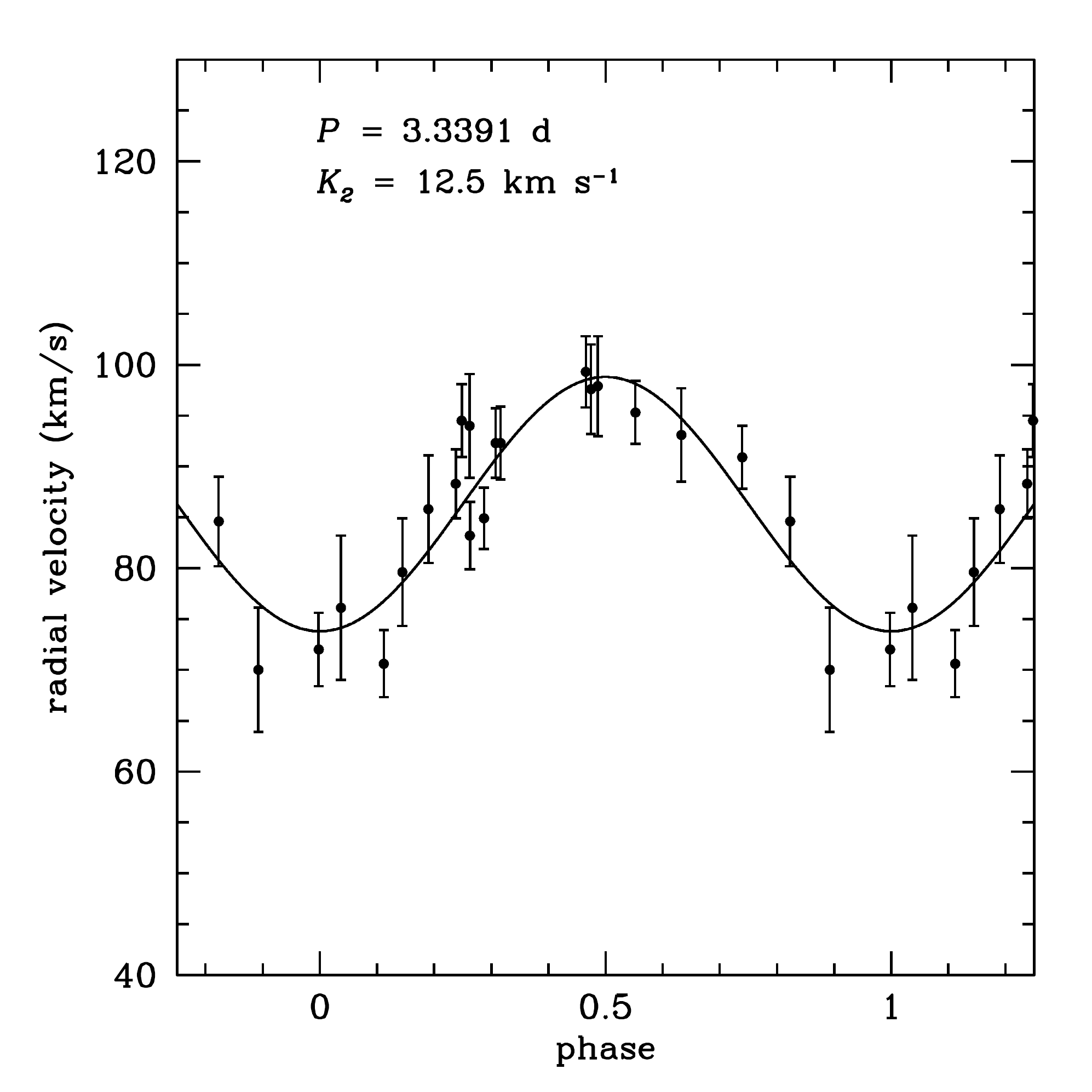}\label{fig:rv}}
  \hfill
  \caption{Radial velocity curve of the red straggler in M10-VLA1, with the best-fit circular Keplerian model overplotted. }
\end{figure}

For an eccentric model, the posterior distribution of the eccentricity is very broad, strongly disfavoring only high eccentricities $> 0.4$. As expected, allowing an eccentric fit slightly improves the model, with a small reduction of the rms from 3.9 to 3.7 km s$^{-1}$ for eccentricities in the range $\sim 0.1$--0.2. These fits still yield periods and semi-amplitudes very similar to the circular case and hence do not affect any of our scientific conclusions. A modest eccentricity is possible, but given the long period and low semi-amplitude, it is not well-constrained with the current set of observations, and we restrict our discussion to the circular case.

The systemic velocity is somewhat surprising: the velocity of M10 itself is 74 km s$^{-1}$, with a central velocity dispersion of about 5 km s$^{-1}$ (Bellazzini et al.~2012; Carretta et al.~2009). Hence M10-VLA1 has a relative velocity close to escape velocity for the cluster. To investigate this possible velocity discrepancy, we used a Besan\c con Galactic model (Robin et al. 2003) to simulate the field star population in a square degree around M10. Of field stars with colors and magnitudes comparable to M10-VLA1, only about 1.3\% had radial velocities as high as that inferred for the systematic velocity of M10-VLA1. Hence, given the radial velocity of this binary and its position very close to the center of M10, we conclude that it is much more likely to be a cluster member than an interloping field star. A future proper motion measurement with the \emph{Hubble Space Telescope} or \emph{Gaia} can confirm this definitively. It is possible that the binary had a recent encounter with another system and received a kick.

If mass transfer is occurring (\S 3.3), the multi-day period implies a low-density, evolved donor, as expected on the basis of the location of the star in the color-magnitude diagram (Figure 4). 

\subsection{Masses}

The semi-amplitude and period together give the mass function $f(M)$

\begin{equation}
f(M) = \frac{P K_{2}^{3}}{2 \pi G} = \frac{M_1 \, (\textrm{sin}\, i)^{3}}{(1+q)^{2}}
\end{equation}

\noindent
where $i$ is the inclination and $q$ the mass ratio $M_2/M_1$. We find $f(M) = 6.7_{-2.1}^{+2.7} \times 10^{-4} M_{\odot}$ using the posterior samples from \S 3.1. There are two possible cases: first, if the visible star is indeed the donor, then the low $f(M)$ immediately implies that the system must be close to face-on. 

To quantify this, we note that the donor mass $M_2$ must be in the range $0.3 M_{\odot} \lesssim M_2 \lesssim 0.8 M_{\odot}$. The upper limit of $0.8 M_{\odot}$ corresponds to the main sequence turnoff mass and $0.3 M_{\odot}$ corresponds to the mass of a heavily-stripped star that still appears to be a red giant in the color-magnitude diagram. As a comparison, the stripped optical companion to a millisecond pulsar in NGC 6397 (COM J1740-5340) has a dynamical mass of 0.22$M_{\odot}$ -- 0.32$M_{\odot}$ and a red color, but is fainter than the main sequence turnoff of the cluster (Ferraro et al.~2003; Mucciarelli  et al.~2013). The red giant optical companion of M10-VLA1 is more luminous than this star, consistent with a comparable or higher mass than COM J1740-5340.

Given the value of $f(M)$, an assumed value of $M_2$ then gives a relationship between the primary mass $M_1$ and the inclination $i$. For the case where the visible star is the secondary, the maximum inclination would be at the extreme case where the primary is a low-mass He white dwarf or a main sequence star with a mass slightly above $\sim 0.3 M_{\odot}$; in this case $i \sim12^{\circ}$. For any less extreme set of assumptions, the inclination would be even lower. If the red straggler is the secondary, {\it M10-VLA1 must be essentially face-on.}

The alternative case is if the red straggler is the source of the X-ray and radio emission.  It might still be the donor if accretion is occurring onto a $\sim 0.2 M_{\odot}$ He white dwarf (see \S 4.4), or it could be the sole source of the X-ray and radio emission, due to chromospheric activity rather than mass transfer. Depending on the red straggler mass, in the former case we find $i = 16-26^{\circ}$, which is still relatively face-on. If no accretion is occurring and the secondary is not a compact object, a wide range of inclinations are allowed. For for a typical case with $M_1 = 0.8 M_{\odot}$ and an inclination of $i = 60^{\circ}$, $M_2 \sim 0.09 M_{\odot}$, near the border between brown dwarfs and the lowest-mass main sequence stars.

If the red straggler is the donor in the binary, an independent estimate of the mass is possible by assuming the star fills its Roche Lobe and using the optical photometry. From the orbital period, the density of the red straggler is $\sim 0.017$ g cm$^{-3}$ (Eggleton 1983). Using the temperature and bolometric luminosity determined above, we find a donor mass of $0.34 M_{\odot}$, which would be consistent with a scenario in the star was substantially stripped. However, we emphasize this mass estimate is strongly dependent on the assumptions that the red straggler is Roche lobe-filling and on the temperature and luminosity used.

\begin{deluxetable}{crr}

\tablecaption{Barycentric Radial Velocities of M10-VLA1 \label{tab:bigg}}
\tablehead{BJD & RV & Error \\
                   (d)  & (km s$^{-1}$) & (km s$^{-1}$) }

\startdata
2457166.6664853 & 83.2 & 3.3 \\
2457166.8171978 & 92.3 & 3.4 \\
2457166.8458049 & 92.3 & 3.6 \\
2457170.6829383 & 99.3 & 3.5 \\
2457170.7138864 & 97.6 & 4.4 \\
2457186.6184706 & 88.3 & 3.4 \\
2457186.7838743 & 84.9 & 3.0 \\
2457196.7163945 & 94.0 & 5.1 \\
2457252.5998639 & 72.0 & 3.6 \\
2457257.5697872 & 97.9 & 4.9 \\
2457473.8156011 & 94.5 & 3.6 \\
2457476.8085067 & 79.6 & 5.3 \\
2457484.8487222 & 95.3 & 3.1 \\
2457508.8464624 & 90.9 & 3.1 \\
2457598.6466511 & 93.1 & 4.6 \\
2457602.6214264 & 84.6 & 4.4 \\
2457603.5856243 & 70.6 & 3.3 \\
2457509.8408503 & 76.1 & 7.1 \\
2457629.5657976 & 70.0 & 6.1 \\
2457630.5586368 & 85.8 & 5.3 \\
\enddata

\end{deluxetable}

\subsection{Optical Spectrum and Emission}

The optical spectra of M10-VLA1 are consistent with a G-type star, with relatively modest metal lines, as would be expected for a star in M10 (with [Fe/H] $\sim -1.5$, Haynes et al.~2008). The most notable feature of the spectra is the presence of H$\alpha$ in emission, which is observed at all six epochs (spanning 16 months) for which spectra covering this region were obtained. By fitting a Gaussian convolved with the instrumental resolution (79 km s$^{-1}$) to rectified spectra in this region, we find that the mean full-width at half-maximum (FWHM) of the H$\alpha$ line is $180\pm25$ km s$^{-1}$ (with the uncertainty representing the standard error of the mean).

In some of the H$\alpha$ spectra, the line appears double-peaked rather than simply broadened, though this is challenging to confirm in individual exposures given the modest width of the line. In Figure 7 we show an average spectrum of six 15-min exposures taken on 2015 May 15, which represents the highest signal-to-noise spectrum of the region. Here the double-peaked nature of the line is obvious. Fitting a single Gaussian model as above yields a FWHM wider than the mean value ($243\pm19$ km s$^{-1}$). We also fit a double-Gaussian model, yielding a velocity difference of the two emission peaks of $147\pm11$ km s$^{-1}$.

\begin{figure}[t]
  \centering
  
  {\includegraphics[width=0.425\textwidth]{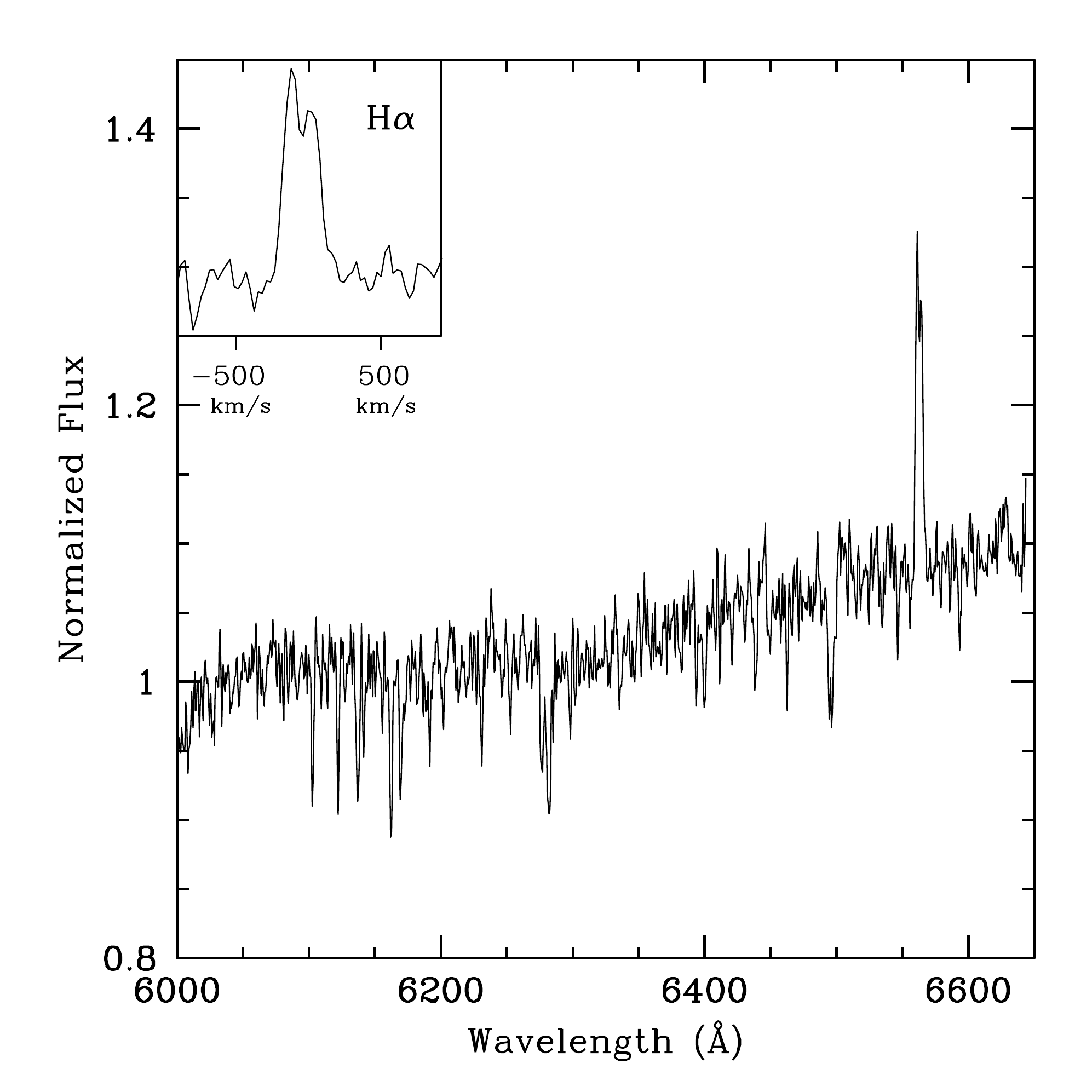}\label{fig:ospec}}
  \hfill
  \caption{Optical spectrum of M10-VLA1, taken on 2015 May 15, over the wavelength range 6000 to 6650 \AA. The inset panel shows a zoom-in on H$\alpha$, where the double-peaked nature of the line is obvious.}
\end{figure}

Given the modest H$\alpha$ velocities and that these values are not orbital averages, we do not wish to over-interpret them. Instead we simply remark that the ratio of the peak separation to the FWHM is $0.60\pm0.06$, exactly the value observed for typical accretion disks around compact objects (e.g., Casares 2015; Casares 2016). Therefore these observations provide evidence that an accretion disk could be present in M10-VLA1.

\section{Discussion}

The X-ray and radio emission and optical photometry and spectroscopy all point to the identification of M10-VLA1 as an interacting binary in the globular cluster M10. While we have observed a red straggler star as one member of this system, the nature of its companion is not clear. We discuss the options in turn.

\subsection{Black Hole}

There are several pieces of observational evidence that suggest a black hole origin for M10-VLA1. First, its ratio of radio to X-ray luminosity puts M10-VLA1 within the scatter of the $L_X$--$L_R$ correlation for quiescent black hole systems (Figure 8; Gallo et al. 2014). The radio continuum spectral index is $\alpha = 1.3\pm0.9$, poorly constrained but consistent with the flat-to-inverted value expected for self-absorbed synchrotron emission from a compact jet, as observed for low-luminosity accreting black holes (e.g., Gallo et al.~2005). We must add that the radio and X-ray observations of M10-VLA1 were not obtained simultaneously, and interpret its position on the $L_X$--$L_R$ diagram cautiously. The Feb 20 2014 VLA observed radio luminosity is shown with both the \emph{Chandra} X-ray detection from May 08 2015 as well as the almost simultaneous Feb 21 2014 \emph{Swift} observation upper limit in Figure 8.  Both points lie above the $L_X$--$L_R$ relation, with the Feb 2014 datapoint somewhat closer the the parameter space occupied by accreting neutron star systems. We also remind the reader that the source is variable in the radio, and multiple simultaneous X-ray and radio observations are needed in the future to determine its relationship with the $L_X$--$L_R$ relation with more certainty.

Another line of argument uses the X-ray luminosity and orbital period of the system. M10-VLA1 has a low X-ray luminosity of $\sim 10^{31}$ erg s$^{-1}$. In the context of an X-ray binary, the X-ray luminosity is  determined by the accretion rate and radiative efficiency of the accretion flow. For an evolved Roche lobe-filling donor, the mass loss rate is thought to be set by the nuclear evolution of the star. The physical mechanism that determines the quiescent radiative efficiency is not well-understood, but it has been argued that black hole binaries typically have lower X-ray luminosities than neutron star X-ray binaries at the same orbital periods (and thus accretion rates), perhaps because accretion luminosity can be advected across the event horizon in a black hole (Garcia et al.~2001). It has also been suggested that black holes simply transfer more energy to jets rather than into the area of the hot accretion disk as hard X-rays (Fender et al.~2003).

\begin{figure*}[t]
\centering
  
  \includegraphics[width=0.8\textwidth]{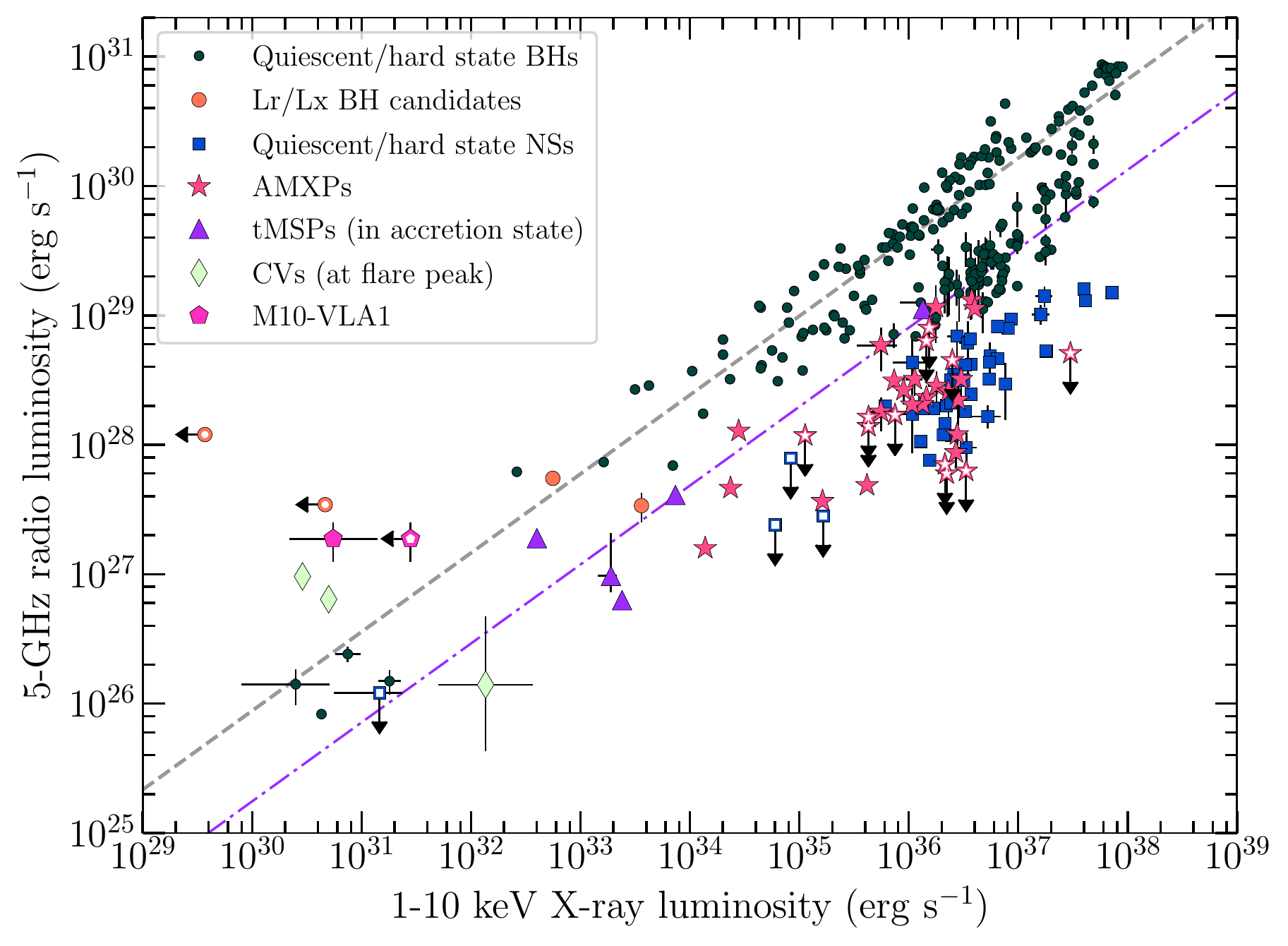}\label{fig:lrlx}
  \hfill
  \caption{The radio/X-ray correlation for accreting compact objects, showing that M10-VLA1 has properties consistent with a quiescent stellar-mass black hole. Hollow points indicate an upper limit. Two magenta pentagons are shown for M10-VLA1, both using the radio luminosity of the 20 Feb 2014 VLA observation with the \emph{Chandra} X-ray detection on 08 May 2015 and \emph{Swift} upper limit of 21 Feb 2014.The dark green circles show known quiescent black holes in the field (Miller-Jones et al. 2011; Gallo, Miller \& Fender 2012; Ratti et al. 2012; Corbel et al. 2013; Rushton et al. 2016; Plotkin et al. 2017). The dotted black line shows the best-fitting $L_R$--$L_X$ correlation for black holes from Gallo et al.~(2014). Orange circles are radio-selected black hole candidates, (Strader et al.~2012; Chomiuk et al.~2013; Miller-Jones et al.~2015; Tetarenko et al.\ 2016b; Bahramian et al.~2017). The purple triangles are transitional millisecond pulsars (Hill et al.~2011; Papitto et al.~2013; Deller et al.~2015; Bogdanov et al. 2017) and the dotted purple line shows their proposed $L_{R}$--$L_{X}$ track. Blue squares are NSs in the hard state, and pink stars are accretion-powered millisecond X-ray pulsars (Migliari \& Fender 2006; Tudor et al. 2017). The light green diamonds are the bright CVs AE Aqr ($L_{X} = 5.0\times 10^{30}$ erg s$^{-1}$), SS Cyg (in outburst; $L_{X} = 1.4\times 10^{32}$ erg s$^{-1}$; Russell et al. 2016), and white dwarf ``pulsar" AR Sco ($L_{X} = 2.9\times 10^{30}$ erg s$^{-1}$; Marsh et al. 2016). Figure adapted from Bahramian et al~(2017).}
\end{figure*}

Using the Reynolds \& Miller (2011) relation for X-ray luminosity vs.~orbital period for quiescent black holes and neutron stars, M10-VLA1, with an orbital period of 3.339 d, has  properties much more consistent with a black hole than a neutron star: a typical neutron star with this period should have  $L_X \gtrsim 5 \times 10^{32}$ erg s$^{-1}$. It is worth noting that some sources ``disobey" this relation: e.g., GS 1354-64, which is much more X-ray luminous than other black holes with comparable orbital periods (Reynolds \& Miller 2011). This is unsurprising given the tenuous physical basis for understanding the radiative efficiency in these systems. Nonetheless, it is still true that the X-ray luminosity of M10-VLA1 would be remarkably low for an accreting neutron star.

Circumstantial evidence for a black hole primary is the possible presence of an accretion disk in the system, as suggested by the unusual donor (consistent with mass transfer), double-peaked H$\alpha$ emission and UV/optical  variability. An accretion disk suggests a compact primary, and as this and the other subsections show, in this case the radio and X-ray emission strongly favor a black hole.

The counterargument is straightforward: if the primary is a black hole, the  inclination must be very low. Even in the extreme case of a $3 M_{\odot}$ black hole, $i = 3.9\pm0.5^{\circ}$. If we instead assume a uniform distribution of black hole masses between 3 and 15 $M_{\odot}$, we find $i = (2.5^{+0.7}_{-0.4})^{\circ}$. Such a low inclination is  unlikely to occur by chance ($\sim 0.1$ $\%$ chance), and in Sec.~4.5 we discuss the possibility that face-on systems would be preferentially observed.

One fact not strongly on either side is the variability of the radio source. There is substantial evidence that the flat-spectrum radio emission from compact jets from neutron stars and black holes is variable on a range of timescales (we discuss white dwarfs below in Sec.~3.3). For example, Miller-Jones et al.~(2008) show that the stellar-mass black hole V404 Cyg has factor of $\sim 3$ variations in its quiescent 8.4 GHz radio continuum flux density on timescales of $< 1$ hr, an observation borne out by a much larger sample of VLA data obtained over decades (R.~Plotkin et al.\ 2018, in preparation). Similar variability persists on longer timescales; the low-luminosity black hole X-ray binary A0620-00 has been seen to vary by a factor of $\sim$2.5 between 2005 and 2013 in quiescence (Din\c{c}er et al.\ 2017).

Finally, we note that Ivanova et al.\ (2017) have recently proposed that red straggler companions (like that of M10-VLA1) are expected for some stellar-mass BHs in globular clusters. The binary is formed in a glancing tidal capture between a BH and a subgiant star, and the interaction strips a few $\times 0.1 M_{\odot}$ off the subgiant. As the donor evolves, it may spend 0.5--1 Gyr in the red straggler portion of the color-magnitude diagram. The long timespan that stars spend as subgiants and their enhanced cross-sections as they swell favor such interactions for subgiants above that of main sequence stars or normal giants (Ivanova et al.~2017). This scenario is consistent with the location of M10-VLA1 near the cluster center, at a projected radius of only 0.2 core radii.

\subsection{RS CVn}

RS CVn systems are detached binary systems with an evolved primary (typically a F/G subgiant or a K giant) and a non-degenerate secondary. Generally the secondary is of similar mass
(Gunn 1996), though here we also consider systems with a wider range of mass ratios. The evolved star shows enhanced chromospheric activity, resulting in variability at a wide range of wavelengths. Relevant for M10-VLA1, RS CVn stars show increased radio and X-ray luminosity compared to similar stars without binary companions (Montesinos et al.~1988; Gunn 1996). 
The orbital periods of RS CVn systems are typically between 2 and 14 days, and shorter period systems are more active due to enhanced tidal synchronization. Activity
in RS CVn stars is manifested in flares, lasting minutes to hours, with order-of-magnitude increases in the X-ray, UV, and radio luminosities (Osten et al.~2000).

RS CVns show non-thermal radio emission associated with the enhanced magnetic field of the rapidly rotating evolved star (Osten et al. 2000; Garc{\'{\i}}a-S{\'a}nchez et al. 2003).  At cm wavelengths, the spectral luminosity typically ranges from 10$^{23}$--10$^{26}$ erg s$^{-1}$ GHz$^{-1}$ (Morris \& Mutel 1988; Drake, Simon \& Linsky ~1989, 1992), and extreme flaring events may reach 10$^{27}$ erg s$^{-1}$ GHz$^{-1}$ (Mutel et al.~1987). The spectral indices of these systems are typically between --1 and 1, with quiescent radio emission flat to steep ($\alpha \lesssim 0$), tending toward inverted ($\alpha \sim 1$) during luminous flares (Gibson et al.~1975; Owen \& Gibson 1978; Mutel et al.~1987). Quiescent radio emission from RS CVn systems shows moderate circular polarization at frequencies above 5 GHz,  but less than 3$\%$ during flares. Flares also produce X-ray emission, with typical luminosities of  $L_X \sim$ 10$^{29}$--10$^{32}$ erg s$^{-1}$. The X-ray and radio emission during flares is correlated (Osten et al.~2000). Standard chromospheric emission lines (Ca H+K; Balmer lines) are commonly observed.

These systems have several overlapping characteristics with those of M10-VLA1. The X-ray luminosity of M10-VLA1 is well within the range of known X-ray luminosities of RS CVn binaries, and the optical counterpart is evolved, as expected for an RS CVn system. The radio spectral luminosity of M10-VLA is at the upper edge of those observed for RS CVn systems in flares, though the low flux density means that any polarization constraints are not useful.

The aspects of the system less consistent with standard RS CVn binaries are the low mass of the secondary for reasonable inclinations and the tentative evidence for an accretion disk (double-peaked H$\alpha$). Double-peaked H$\alpha$ has been exceptionally observed in RS CVn systems, e.g., in SZ Pis (Bopp 1981), where it was attributed to circumstellar material ejected in transient mass transfer events. Unusually for RS CVn stars, the subgiant in SZ Pis nearly fills its Roche lobe. This suggests that if the origin of the H$\alpha$ is similar in M10-VLA1, then it is also likely to be close to Roche lobe-filling. The analysis in \S 3.2 would then imply that the subgiant in M10-VLA1 is likely to be a stripped, low-mass star.

We note that the X-ray luminosity and orbital period of M10-VLA1 are consistent with those of some binaries containing sub-subgiant or red straggler primaries (Geller et al.~2017), though such systems do not typically have evidence for accretion, and their radio continuum properties are unknown. At least some sub-subgiants are found in compact binaries (e.g., Mucciarelli et al.~2013).

\subsection{Neutron Star}

Compact binaries with neutron star primaries are frequently identified in globular clusters. Many of these neutron stars are millisecond radio pulsars spun up by accretion in a dynamically-formed binary, but for which the accretion has temporarily or permanently ceased. The flat to inverted spectrum radio continuum emission and evidence for an accretion disk from M10-VLA1 are unlike the steep spectrum radio emission and lack of accretion observed for normal millisecond pulsars (e.g., Kramer et al. 1999).

As mentioned above in Sec.~4.1,  M10-VLA1 does not have properties consistent with being an actively accreting neutron star X-ray binary: its X-ray luminosity is at least a factor of $\sim 40$ lower than expected for an accreting neutron star at its orbital period, and it is much more radio bright than would be expected (Figure 8). Indeed, the only class of neutron stars with detected flat-spectrum radio emission at X-ray luminosities $< 10^{34}$ erg s$^{-1}$ are the transitional millisecond pulsars that switch between accretion-powered disk states and rotation-powered pulsar states on timescales of days to years (Archibald et al. 2009; Bond et al. 2002; Hill et al. 2011; Papitto et al. 2013; Bassa et al. 2014; Deller et al. 2015). During their disk state, transitional millisecond pulsars emit flat spectrum radio emission, which is typically interpreted as compact radio jets. They may also show double peaked H${\alpha}$ emission originating from an accretion disk.

While our radio detection was not simultaneous with the \emph{Chandra} X-ray detection, we can constrain this scenario through the quasi-simultaneous \emph{Swift} observations, which limit an X-ray source at this position to $< 5.2 \times 10^{31}$ erg s$^{-1}$ (0.5--10 keV). This is a factor of 10 below the X-ray luminosities observed for transitional millisecond pulsars in even their ``low mode" disk states (de Martino et al. 2013; Patruno et al. 2014; Bogdanov et al. 2017). In fact, the seven \emph{Swift} observations all have flux limits at least a factor of 5 below the typical $2 \times 10^{33}$ erg s$^{-1}$ X-ray 
luminosity observed for accreting transitional millisecond pulsars (a mixture of the ``low" and ``high" modes),
and some \emph{Swift} observations are much deeper. 

The ratio of radio to X-ray flux of M10-VLA1 is also dissimilar to the average values for known transitional millisecond pulsars. Bogdanov et al.~(2017) show that the transitional PSR J1023+0023 shows anti-correlated radio and X-ray
variability in the disk state, including periods in which the system can enter the parameter space in $L_X$--$L_R$
occupied by accreting black holes. However, the \emph{Swift} upper limit for M10-VLA1 is still about a magnitude fainter than would be expected for a transitional millisecond pulsar in the low mode disk state at the observed radio luminosity.
M10-VLA1 also lacks the short-term ($<$ 2 hours) radio variability observed in the PSR J1023+0023 disk state.

Overall, the radio and X-ray evidence suggests that M10-VLA1 does not have properties similar to known
transitional millisecond pulsars in their disk states, though we cannot definitely rule out the possibility
that we are observing such a system in an ``sub-subluminous" disk state not yet observed among
the confirmed transitional millisecond pulsars.

Considering the dynamical evidence, if we assume a neutron star in the mass range 1.4--2.0 $M_{\odot}$, the inclination inferred is $i = 5.1\pm0.7^{\circ}$. Hence this scenario has the same drawback as the black hole case (an unlikely face-on inclination) and the additional issue that the properties of M10-VLA1 are inconsistent with those of known low-mass X-ray binaries containing neutron stars. Hence a neutron star primary is disfavored, though not ruled out.

\subsection{White Dwarf}

Here we discuss a number of possibilities in which the red straggler is in a binary with a white dwarf.

\subsubsection{Flare from Accreting White Dwarf}

Miller-Jones et al.~(2015) have exhaustively catalogued evidence for bright radio flares among accreting white dwarfs in the context of interpreting the stellar-mass black hole candidate X9 in the globular cluster 47 Tuc, and we do not repeat their discussion here. In brief, while accreting white dwarfs do emit variable radio continuum emission, this emission is generally 1--4 orders of magnitude fainter than observed for M10-VLA1 (e.g., Coppejans et al.~2015, 2016). A few well-studied intermediate polars (AE Aqr, DQ Her) occasionally emit bright flares, but these are still typically less luminous than M10-VLA1 or decay on short ($<$ hr) timescales (Bastian et al.~1988; Abada-Simon et al.~1993; Pavelin et al.~1994). A bright, very brief ($< 20$ min) 15 GHz flare was observed from the dwarf nova SS Cyg (Mooley et al. 2017), which would have a scaled flux density of 10--13 $\mu$Jy at the distance of M10.

As discussed in \S 3.2, a white dwarf accretor also does not easily explain the low semi-amplitude of the binary. The most favorable case would be if the white dwarf is a low-mass $\sim 0.2 M_{\odot}$ He white dwarf formed through mass transfer during the evolution of the initially more massive star, and we are now witnessing the evolution of the initially less massive (but now more massive) star. The evolving red straggler would likely be more massive than the white dwarf, so by standard criteria stable mass transfer is unlikely. Some recent theoretical work suggests a wider range of mass ratios might allow stable mass transfer from giants (Pavlovskii \& Ivanova 2015), which could allow accretion onto a low-mass white dwarf. 

Nonetheless, even if we take their most extreme case of a donor to accretor mass ratio of 2.2, a relatively face-on inclination of $i \lesssim 19^{\circ}$ is still required, and this source would still be the most radio-loud accreting white dwarf known. In addition, while we do observe optical/UV variability from the system, there is no strong evidence for a substantial UV excess indicative of a hot disk, though the \emph{Swift}/UVOT absolute UV limit on an outburst at the time of the radio detection ($UVW2_0 > 2.9$) is not strong. As a comparison, the variable, long-period accreting white dwarf AKO 9 in the globular cluster 47 Tuc has an inferred $UVW2_0 \sim 3.8$--4.3 (Edmonds et al.~2003; Knigge et al.~2003).

\subsubsection{A White Dwarf ``Pulsar"}

The white dwarf binary AR Sco was previously classified as an intermediate polar, but has has recently been shown to emit bright, pulsed radio continuum radiation, and it may represent the first member of a separate class of ``white dwarf pulsars". The radio emission appears to originate in an interaction between a close magnetic white dwarf--M dwarf binary rather than primarily from accretion. AR Sco has a 0.15 d period (Marsh et al.~2016; Buckley et al.~2016), quite unlike the long 3.339 d period of M10-VLA1. If the radio emission in M10-VLA1 is not associated with accretion, the double-peaked H$\alpha$ line is difficult to explain, and this scenario has the same requirement of a face-on inclination as a normal accreting white dwarf. Therefore, it seems reasonable to argue that this scenario is disfavored, but a stronger statement would require a better understanding of the white dwarf pulsar mechanism. It is worth pointing out that AR Sco is very nearby (116 pc; Marsh et al.~2016) and so white dwarf pulsars might well be common.

\subsubsection{An Ionized Red Straggler Wind}

A third scenario posits an enhanced wind from the red straggler, such as for a symbiotic system, with the X-rays produced by accretion and the radio emission due to thermal radiation from the ionized wind. The X-ray luminosity suggests an accretion rate of $\sim 10^{-10}-10^{-11} M_{\odot}$ yr$^{-1}$ (assuming a boundary layer on a 0.2 M$_{\odot}$ white dwarf; Kuulkers et al.\ 2006). Assuming the radio emission is from an ionized wind with a velocity of 100 km s$^{-1}$, this mass loss rate would be undetectable in the radio (Panagia \& Felli 1975); a much higher mass loss rate of $\sim 10^{-8} M_{\odot}$ yr$^{-1}$ would be necessary to produce a thermal radio source with $\sim 10\mu$Jy. Such a mass loss rate is much higher than observed for metal-poor giants of this luminosity (Dupree et al.~2009). The natural expectation is that this thermal emission would also not be variable; significant variation in the ionizing source would be necessary to produce the time-variable radio emission observed. Hence we believe this scenario is unlikely.

\subsubsection{The Red Straggler Alone}

A final option is that, if the white dwarf is \emph{not} accreting, the X-ray and radio emission could be associated entirely with the rapidly rotating red straggler, as for the RS CVn scenario. As discussed in \S 3.2, the inclination requirements for this scenario are less extreme than for more massive white dwarfs, neutron stars, or black holes, but a face-on inclination with $i \lesssim 26^{\circ}$ is still necessary.

\subsection{A Face-On Binary: Relativistic Beaming?}

If the faint companion in M10-VLA1 is a compact object, the binary's radial velocity curve implies that it must be relatively face-on. Face-on inclinations are a priori unlikely. For example, given a random distribution of orientations in cos($i$), the value $i < 11.1^{\circ}$ (as implied for a compact object primary) would occur by chance only 1.9\% of the time. This leads us to consider whether any selection biases might exist for radio or X-ray emission in favor of face-on systems. 
Tentative evidence for a similar bias toward face-on low-mass X-ray binaries has been observed at $\gamma$-ray wavelengths in systems with lower X-ray luminosities (Britt et al.~2017), including the candidate transitional millisecond pulsar
3FGL J1544.6--1125, which has an inferred inclination of 5--8$^{\circ}$.

There are a limited number of physical mechanisms that would lead to a bias for face-on systems.  Perhaps the most promising candidate is relativistic beaming of the radio emission. The X-ray emission might also be beamed depending on its origin.

For $\beta = v/c$, the jet Lorentz factor is $\Gamma = (1-\beta^2)^{-1/2}$. Assuming a flat radio spectral index $\alpha=0$ and a continuous jet, the observed flux density is boosted by a factor $[\Gamma (1-\beta \, cos \, i )]^{-2} + [\Gamma (1+\beta \, cos \, i)]^{-2}$, where the second term is negligible for face-on inclinations.

As a proof of concept, we consider the case where the X-ray flux is not beamed and the radio flux of M10-VLA1 is consistent with the black hole $L_X$--$L_R$ relation in the rest frame. This would require a beaming factor of $\sim 5$. For any inclination allowed by the dynamical analysis (\S 3.2), this beaming would require $\Gamma \sim 1.32$--1.36, corresponding to $\beta \sim$ 2/3, for the inferred flux boost. Beaming factors of 10 or even 20 are easily reached with $\beta \lesssim 0.9$, still in the mildly relativistic regime. The beaming required to move a source from the (admittedly poorly-defined) $L_X$--$L_R$ relation
for transitional millisecond pulsars to the location of M10-VLA1 is a factor of $\sim 4$--10 higher than the black hole case.

For relativistic beaming to be a likely explanation for M10-VLA1, there are two requirements for the source class. First, it requires a significant population of sources with flux densities below our detection limit but which become detectable if the orientation is favorable. Second, it requires that these sources regularly produce at least mildly relativistic jets in the quiescent regime.

Both requirements would tend to favor neutron star or black hole binaries: while accreting white dwarfs might well be common in globular clusters, those that emit bright radio flares are unusual, and to our knowledge relativistic jets have not been proposed to exist for dwarf nova systems outside of outburst.

Very little is known definitively about the Lorentz factors of jets in X-ray binaries, especially for low-luminosity sources with $L_X < 10^{34}$ erg s$^{-1}$. Heinz \& Merloni (2004) show that the moderate scatter in the $L_X$--$L_R$ correlation for black holes in the low/hard state implies that the width of the Lorentz factor distribution should be relatively small, but also that no upper limit on $\Gamma$ can be derived from this correlation. 

In a study of the stellar-mass black hole GX 339-4 undertaken in the hard state (with $L_X \gtrsim 10^{36}$ erg s$^{-1}$), Casella et al.~(2010) show that $\Gamma \gtrsim 2$ near the jet base. Gallo et al.~(2014) considered whether the different normalizations of $L_X$--$L_R$ for different stellar-mass black holes could be explained by beaming, with no compelling evidence that this is the case. Russell et al.~(2015) argue that the steep $L_X$--$L_R$ relation of the radio-bright face-on ($i = 4$--15$^{\circ}$; Russell et al.~2014) candidate stellar-mass black hole MAXI J1836-194 could potentially be explained by a decreasing Lorentz factor as the luminosity decayed to below $10^{36}$ erg s$^{-1}$ after a failed outburst, but such an argument cannot be universally applied for stellar-mass black holes (Heinz \& Merloni 2004; Soleri \& Fender 2011). 

Overall, we conclude that there is no strong evidence for or against the presence of relativistic jets among low-luminosity X-ray binaries. If such binaries \emph{do} generally host mildly relativistic jets that could result in flux boosts from beaming, then we would expect the discovery of other face-on systems, an idea readily testable with future observations.

\section{Summary and Future Work}

The central result of the paper is the discovery of a interacting binary star with an evolved red straggler companion in the globular cluster M10. The identity of this star's companion is uncertain.

As discussed in \S 4.1, the observed properties of the system are all consistent with a black hole primary, with the exception of the low mass function, which could be explained if M10-VLA1 is face-on and there is a selection effect favoring radio detection of face-on systems. \S 4.2 discussed the alternative possibility that M10-VLA1 is an extreme flaring RS CVn binary, which would better explain the dynamical data, but is in tension with the long-term evidence for accretion in the binary.

There are a number of avenues for future work that could help distinguish between these possibilities. For a nominal $10 M_{\odot}$ black hole primary, the semi-major axis would be $\sim 20 \mu$as, which might be marginally detectable with \emph{Gaia} astrometry given the brightness of the red straggler (Barstow et al. 2014). Perhaps more attainable, a simultaneous UV to near-IR spectral energy distribution, well-sampled in orbital phase, would allow improved modeling of the system.  Simultaneous deep radio and X-ray observations, taken over multiple epochs, would help distinguish among classes of accreting compact objects. Theoretical modeling of the evolution of the system, as done for some specific sub-subgiant systems by Leiner et al.~(2017), would also be desirable. 

In any case, it is clear that radio continuum imaging offers unique insights on the close binary populations in Galactic globular clusters.

\acknowledgments

We thank an anonymous referee for many helpful comments that improved the paper.
We also thank Mike Siegel for help with the \emph{Swift}/UVOT analysis. We acknowledge support from NSF grant AST-1308124 and a Packard Fellowship. Partial support for this work was provided by NASA through Chandra Award Number GO5-16036X issued by the Chandra X-ray Observatory Center, which is operated by the Smithsonian Astrophysical Observatory for and on behalf of NASA under contract NAS8-03060. This material is also based upon work supported by the NSF under Grant No. 1515211. This work was partially supported by NASA Swift grant NNX16AN73G. JCAM-J is the recipient of an Australian Research Council Future Fellowship (FT140101082). C.O.H. and G.R.S. acknowledge support from NSERC Discovery Grants.
 
The National Radio Astronomy Observatory is a facility of the National Science Foundation operated under cooperative agreement by Associated Universities, Inc. We acknowledge the use of public data from the \emph{Swift} data archive. The scientific results reported in this article are based in part on observations made by the Chandra X-ray Observatory. This paper is partially based on observations made with the NASA/ESA Hubble Space Telescope, and obtained from the Hubble Legacy Archive, which is a collaboration between the Space Telescope Science Institute (STScI/NASA), the Space Telescope European Coordinating Facility (ST-ECF/ESA) and the Canadian Astronomy Data Centre (CADC/NRC/CSA). This paper is partially based on observations obtained at the Southern Astrophysical Research (SOAR) telescope, which is a joint project of the Minist\'{e}rio da Ci\^{e}ncia, Tecnologia, e Inova\c{c}\~{a}o (MCTI) da Rep\'{u}blica Federativa do Brasil, the U.S. National Optical Astronomy Observatory (NOAO), the University of North Carolina at Chapel Hill (UNC), and Michigan State University (MSU).

{}

\end{document}